\documentclass[superscriptaddress,twocolumn,prb,floatfix,nobibnotes]{revtex4}

\usepackage{url}
\usepackage{bm,bbm}
\usepackage{graphicx}
\usepackage{color}
\usepackage[colorlinks=true, urlcolor=blue, linkcolor=blue, citecolor=blue, pdftex]{hyperref}
\usepackage[english]{babel}

\usepackage{soul}
\usepackage{blindtext}

\begin{document}

\title{Optimization of infinite projected entangled pair states: The role of multiplets and their breaking}

\author{Juraj Hasik}
\email[]{jhasik@sissa.it}
\affiliation{SISSA-International School for Advanced Studies, Via Bonomea 265, I-34136 Trieste, Italy}
\author{Federico Becca}
\affiliation{Dipartimento di Fisica, Universit\`a di Trieste, Strada Costiera 11, I-34151 Trieste, Italy}

\date{\today}

\begin{abstract}
The infinite projected entangled pair states (iPEPS) technique [J. Jordan {\it et al.}, Phys. Rev. Lett. {\bf 101}, 250602 (2008)] has been widely used in the recent
years to assess the properties of two-dimensional quantum systems, working directly in the thermodynamic limit. This formalism, which is based upon a tensor-network 
representation of the ground-state wave function, has several appealing features, e.g., encoding the so-called area law of entanglement entropy by construction; 
still, the method presents critical issues when dealing with the optimization of tensors, in order to find the best possible approximation to the exact ground state
of a given Hamiltonian. Here, we discuss the obstacles that arise in the optimization by imaginary-time evolution within the so-called simple and full updates and 
connect them to the emergence of a sharp multiplet structure in the ``virtual'' indices of tensors. In this case, a generic choice of the bond dimension $D$ is not 
compatible with the multiplets and leads to a symmetry breaking (e.g., generating a finite magnetic order). In addition, varying the initial guess, different final 
states may be reached, with very large deviations in the magnetization value. In order to exemplify this behavior, we show the results of the $S=1/2$ 
Heisenberg model on an array of coupled ladders, for which a vanishing magnetization below the critical interladder coupling is recovered only for selected values 
of $D$, while a blind optimization with a generic $D$ gives rise to a finite magnetization down to the limit of decoupled ladders.
\end{abstract}

\maketitle

\section{Introduction}\label{sec:intro}

Solving many-body problems in generic two- and three-dimensional spatial dimensions represents a great challenge in modern physics. A less ambitious project is to
obtain accurate representations for ground states of local Hamiltonians. In one spatial dimension, tremendous progress has been achieved by the density-matrix
renormalization group (DMRG) method~\cite{white1992}, which allowed to obtain the correct physical behavior with extraordinary accuracy for several fermionic and
bosonic problems, including the evaluation of the spin gap in the $S=1$ Heisenberg chain~\cite{white1993}. Although the original formulation of this approach was
motivated by a numerical renormalization technique in which a reduced basis set is constructed iteratively to approximate the actual ground state, the modern view 
of DMRG is based upon its underlying variational {\it ansatz}, a so-called  matrix poduct state (MPS)~\cite{ostlund1995,verstraete2008}. Here, the amplitude of the 
wave function for a given configuration of the basis set is represented by a product of matrices, whose linear size is fixed by a parameter $D$, the so-called bond 
dimension. The accuracy of the variational {\it ansatz} can be systematically improved by increasing $D$~\cite{verstraete2006}. Remarkably, the MPS {\it ansatz} can 
be directly applied to an infinite chain~\cite{vidal2007,orus2008}, where it is dubbed iMPS. In practice, the optimization of matrices is performed either by DMRG 
or the time-dependent variational principle~\cite{tdvp}, while a simpler alternative is to use the so-called time-evolved block decimation (TEBD). In one spatial dimension 
(with short-range interactions), very efficient calculations can be performed, working with only a few tensors (one on each site of a given unit cell) embedded in an 
effective environment, which can be easily incorporated without any additional computational effort~\cite{vidal2007,orus2008}. Here, the great simplification comes 
from the fact that the environment, which is instrumental in building reduced density matrices of any subsystem, is given by a tensor product of two vectors of 
dimension $D^2$ (in the auxiliary bond space), corresponding to left and right boundaries of the subsystem.

The reformulation of DMRG into MPS (together with a deep understanding of entanglement properties in quantum systems) has been crucial to define generalizations to 
higher dimensions that go beyond the original extension of DMRG, where a one-dimensional snake-like path is used to cover the entire system~\cite{white1998}. The 
general framework of these approaches goes under the name of tensor networks and is based upon the definition of a variational wave function that is written as a 
(generalized) trace over the product of tensors, which are usually defined on each site of the lattice. In analogy with MPS, each tensor possesses a  physical index
(whose dimension is dictated by the physical Hilbert space) and virtual indices (whose dimension $D$ determines the number of variational parameters). The most 
straightforward extension of MPS is given by projected entangled pair states (PEPS)~\cite{verstraete2004,verstraete2006b,murg2007}. Also PEPS can be embedded to 
consider infinite systems, thus leading to the so-called iPEPS~\cite{jordan2008,orus2009}. 

The major issue of PEPS and iPEPS algorithms is the high computational cost when the bond dimension $D$ is increased. While in one dimension, MPS (or iMPS) can easily
deal with a bond dimension up to $D \approx 10^{4}$, in two dimensions, within PEPS (or iPEPS) we are limited to $D \approx 10$ or $D \approx 25$~\cite{txiang2017}, 
depending on the actual numerical procedure. This is particularly relevant when we search for accurate approximations of the true ground state in highly-entangled 
phases. In principle, this task can be done by performing a discretized imaginary-time evolution~\cite{jordan2008} in which, at every step, we minimize the distance 
between the evolved state, which has an enlarged bond dimension, and a tensor network with the original bond dimension $D$. Within iMPS, this procedure can be 
performed efficiently and exactly (apart from the Trotter-Suzuki discretization error) by imposing Vidal's canonical form, where the environment is given by diagonal
$D$-dimensional matrices (dubbed {\it weights}), sitting between on-site tensors~\cite{vidal2007}. They are modified together with on-site tensors at each step and, 
therefore, provide the exact environment and distance at no additional cost. By contrast, for a generic iPEPS the exact environment is inaccessible and instead an 
approximation must be found at every step of the evolution, in order to evaluate the distance between iPEPS. In this respect, several approaches have been 
proposed, as for example the one that is based upon the so-called corner transfer matrix (CTM) technique~\cite{orus2009}, which was introduced in classical 
statistical physics to accurately approximate partition functions~\cite{nishino1996,nishino1998}. Currently, the computation of the  environment presents the main 
bottleneck of iPEPS, scaling polynomially in $D$, but with a very high power $O(D^{9})$ or $O(D^{10})$ depending on the exact scheme used~\cite{txiang2017}.

Within the simple update (SU) technique~\cite{jiang2008}, the effective environment is severely approximated by a product of weights, as a straightforward 
generalization of the one-dimensional case. Even though the optimization procedure is relatively fast and allows us to reach large values of $D$, typically, it does 
not give the correct description of highly entangled ground states. Therefore, a more refined approach making use of an accurate environment, dubbed the full update (FU), 
was proposed and developed~\cite{jordan2008,phien2015}. Intermediate approaches, which interpolate between SU and FU, have also been suggested~\cite{lubasch2014}. 
Still, in the pursuit of optimization that circumvents the flaws of FU (and  SU), alternative schemes to optimize tensor networks by directly minimizing the 
(variational) energy were recently proposed~\cite{corboz2016,vanderstraeten2016}, including the possibility to utilize algorithmic differentiation to evaluate 
energy derivatives~\cite{liao2019}. While providing the most accurate results in the context of tensor networks for Heisenberg or quantum Ising models~\cite{rader2018}, 
their applicability beyond one-site invariant iPEPS has so far been limited due to their inherent complexity, which leaves FU as the dominant method for simulating 
tensor networks with large unit cells. 

In this paper, we assess the accuracy of iPEPS, defined by a finite bond dimension $D$ and optimized with SU and FU techniques, to describe correctly non-magnetic 
ground states with strong local entanglement, notably the existence of nearest-neighbor singlets. This goes beyond the case of the trivial paramagnetic phase that 
appears in the quantum Ising model~\cite{sachdevbook}, which is adiabatically connected to a product state over each lattice site. 
Indeed, the presence of local 
entanglement induces a nontrivial structure in the virtual space, which is easily broken by a blind opimization, thus leading to some symmetry-breaking mechanism, 
e.g., the generation of magnetic order in the ground-state wave function. This fact has important effects when analyzing a quantum phase transition between magnetically 
ordered and disordered phases, possibly obscuring its nature. In order to illustrate these kinds of issues, we consider the $S=1/2$ Heisenberg model on a set of coupled 
two-leg ladders:
\begin{equation}\label{eq:hamilt}
{\cal H} = J \sum_{R} {\bf S}_{R} \cdot {\bf S}_{R+x} + \sum_{R} J_{R} {\bf S}_{R} \cdot {\bf S}_{R+y},
\end{equation}
where ${\bf S}_{R}=(S^x_{R},S^y_{R},S^z_{R})$ is the $S=1/2$ operator on the site $R=(x,y)$ of a square lattice and $J_{R}=J$ or $J_{R}=\alpha J$, depending on the
parity of $y$. This model interpolates between the Heisenberg model on the square lattice (when $\alpha=1$) and a system of decoupled two-leg ladders (when $\alpha=0$).
In the former case, the ground state has N\'eel antiferromagnetic order and gapless excitations (spin waves); instead, in the latter case, no long-range magnetic order
is present, and the spectrum is fully gapped. Therefore, a quantum phase transition exists at a finite value of the interladder coupling $\alpha$~\cite{sachdev2000},
as detected by using quantum Monte Carlo methods at zero temperature~\cite{matsumoto2001,capriotti2002}. In particular, the precise location of the quantum phase 
transition has been determined with high accuracy, i.e., $\alpha_c=0.31407(5)$, also suggesting that the critical properties are described by the same universality 
class as that of the classical three-dimensional Heisenberg model~\cite{matsumoto2001}.

Our calculations show that the paramagnetic phase that is stable for $\alpha<\alpha_c$ is built by tensors having a particular structure that does not fit with a generic 
value of $D$. As a consequence, the optimization performed within SU or FU schemes generally leads to a symmetry-broken state with a small but finite magnetization. 
The correct vanishing magnetization is obtained only for a few selected values of $D$, making it difficult to perform scaling for $D \to \infty$. Moreover, for generic
$D$, especially in the paramagnetic region, the effective energy landscape appears very rough, featuring many nearly degenerate states with substantially different 
magnetizations. Our results strongly suggest that, within iPEPS (or PEPS), it is extremely important to make use of symmetries in the tensors, as suggested in
Ref.~\cite{jiang2015} and developed in Ref.~\cite{mambrini2016}. 

This paper is organized as follows: in Sec~\ref{sec:methods}, we briefly describe the iPEPS method and its optimization based upon SU and FU; 
in Sec.~\ref{sec:results}, we present the numerical calculations. Finally, in Sec.~\ref{sec:concl}, we draw our conclusions.  

\section{Method}\label{sec:methods}

\begin{figure}[t]
\includegraphics*[width=\columnwidth]{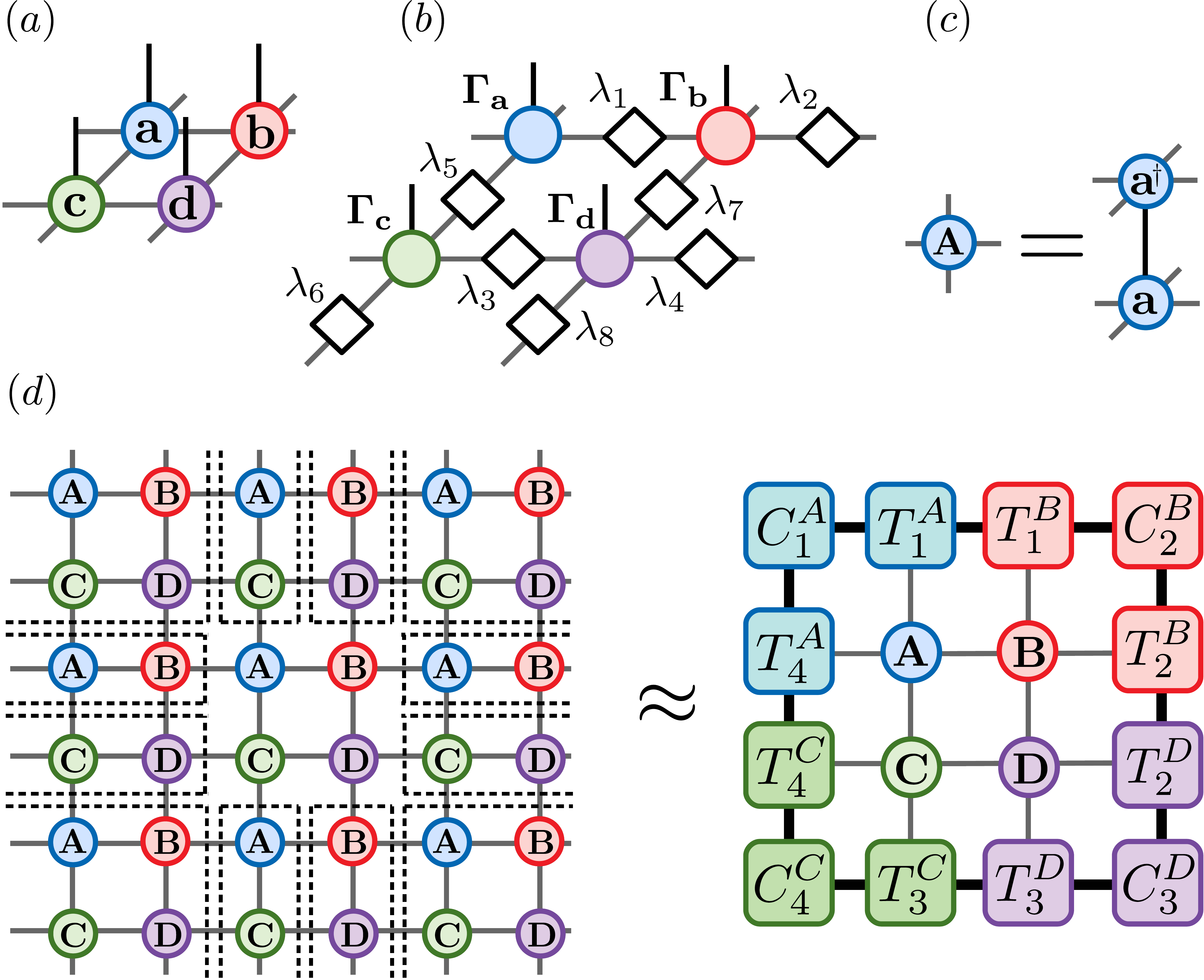}
\caption{\label{fig:ctm}
(a) Regular iPEPS {\it ansatz} for a $2 \times 2$ unit-cell. The horizontal lines represent contracted auxiliary indices of bond dimension $D$, while the vertical lines 
are physical indices. (b) iPEPS {\it ansatz} for a simple update within the same unit cell. (c) Tensor $A$ given by the contraction of on-site tensor $a$ and its complex 
conjugate $a^\dag$ through the physical index. Each index of the double-layer tensor has bond dimension $D^2$. (d) Part of an infinite double-layer tensor network 
corresponding to the norm of an iPEPS given by the ansatz in (c). The dashed lines split the whole network into the central subregion and the environment which is approximated by 
a set of environment tensors $\{ C,T \}$ with bond dimension $\chi$ (bold lines).}
\end{figure}

We parametrize the ground state by an iPEPS $|\Psi(a,b,c,d)\rangle$ with a $2 \times 2$ unit cell containing four different on-site tensors $a$, $b$, $c$, and $d$, 
with auxiliary bond dimension $D$ [see Fig.~\hyperref[fig:ctm]{1(a)}]. Within SU, this {\it ansatz} is ``augmented'' by the inclusion of diagonal matrices $\{ \lambda \}$
(the so-called weights): on each non-equivalent bond between the on-site tensors $\Gamma_a$, $\Gamma_b$, $\Gamma_c$ and $\Gamma_d$, thus leading to a state denoted
as $|\Phi(\Gamma_a,\Gamma_b,\Gamma_c,\Gamma_d,\lambda_1, \dots, \lambda_8)\rangle$ [see Fig.~\hyperref[fig:ctm]{1(b)}]. For the purpose of computing the environment, we can 
absorb the weights into the tensors, e.g., $a=\Gamma_a \sqrt{\lambda_1}\sqrt{\lambda_2}\sqrt{\lambda_5}\sqrt{\lambda_6}$, thus recovering the original form
$|\Psi(a,b,c,d)\rangle$. 

To evaluate observables for a given state, we employ the directional CTM algorithm to construct the environments relative to each site in the $2 \times 2$ unit cell,
as described in Ref.~\cite{corboz2014}. Then, relevant reduced density matrices $\{ \rho \}$ are obtained by combining environments with on-site tensors. Contrary to 
the original version of CTM~\cite{orus2009}, this one leads to gauge-invariant observables for unit cells with more than one tensor. At its core, CTM approximates the 
environment of any spatial subregion of the system by a set of matrices (the so-called corners) $\{ C \}$ of dimension $\chi \times \chi$ and rank-3 tensors (the 
so-called half-row/-column tensors) $\{ T \}$ of dimensions $D^2 \times \chi \times \chi$ [see Fig.~\hyperref[fig:ctm]{1(d)}]; here, for a fixed $D$, the size of $\chi$ governs 
the accuracy of the calculations. 
\begin{figure}[t]
\includegraphics*[width=\columnwidth]{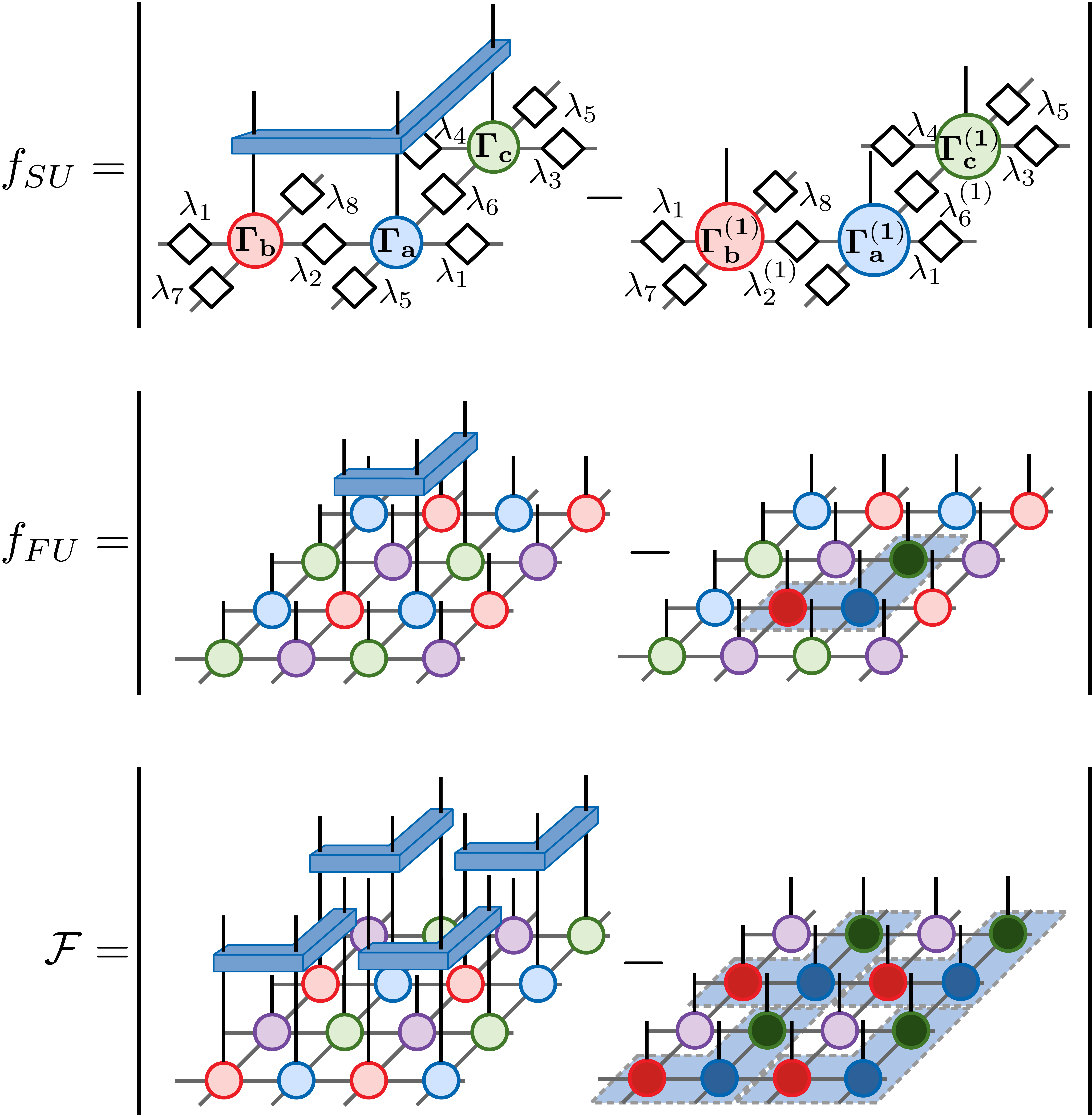}
\caption{\label{fig:distance}
Expressions representing distances between different tensor networks used in SU and FU. Top: Distance to be minimized within the SU procedure given by Frobenius norm of 
the difference between two networks when reshaped into matrices. Middle:  Distance to be minimized in FU between iPEPS $|\Psi_0\rangle$ parametrized by four tensors 
$\{ a^{(0)},b^{(0)},c^{(0)},d^{(0)} \}$ (light-colored circles) with a single three-site Trotter gate $u_{a,b,c}$ applied and the iPEPS $|\Psi_{\rm loc}\rangle$, 
containing three unknown tensors $a^{(1)},b^{(1)},c^{(1)}$ (dark-colored circles) of bond dimension $D$. Bottom: Distance between imaginary-time evolved iPEPS 
$U_{a,b,c}|\Psi_0\rangle$ with the gate $u_{a,b,c}$ applied on {\it all} equivalent triplets $a^{(0)},b^{(0)},c^{(0)}$ and the iPEPS $|\Psi_1\rangle$ 
parametrized by the tensors $\{ a^{(1)},b^{(1)},c^{(1)},d^{(1)}=d^{(0)} \}$.}
\end{figure}
Observables are then recovered as $\langle O \rangle = \lim_{\chi \rightarrow \infty} \text{Tr}[\rho(C,T)O]$. Environment tensors $\{ C,T \}$ 
are defined by the fixed point of an iterative procedure involving contraction of double-layer tensors [see Fig~\ref{fig:ctm}(c)] and truncation of the intermediate 
results (for details see Ref.~\cite{corboz2014}). In practice, the convergence of the observables with $\chi$ is fast, especially for weakly entangled states. 
Nevertheless, CTM is the main computational bottleneck since its complexity  scales as $O((\chi D^2)^3)$ due to the singular value decomposition (SVD) involved in the 
construction of the projectors, which facilitate the truncation.
However, since we need only the leading $\chi$ singular values and corresponding singular vectors to construct the projectors we can use the truncated SVD. 
In this work, we employ two types of truncated SVD algorithms, randomized SVD~\cite{voronin2015} and SVD based on the Arnoldi iteration~\cite{arpack}, reducing the 
computational cost of CTM down to $O(\chi^3 D^4)$.

Optimization of iPEPS is performed by two imaginary-time evolution algorithms (i.e., SU and FU). We consider the case in which the Hamiltonian is the sum of terms 
acting on two sites, as in Eq.~(\ref{eq:hamilt}), namely, ${\cal H}= \sum_{R,R'} h_{R,R'}$. For a sufficiently small imaginary time $\tau$, we simplify the evolution 
operator $U=\exp(-\tau {\cal H})$ by the symmetric Trotter-Suzuki discretization into a product of either two-site or three-site gates, $u_{R,R'} = \exp(-\tau h_{R,R'})$ 
and $u_{R,R',R''} = \exp[-\tau (h_{R,R'}+h_{R',R''})]$, respectively. Within the $2 \times 2$ unit cell, the three sites $R$, $R'$, and $R''$ are arranged to form all 
possible \textsf{L}-shaped patterns covering the two-dimensional lattice. Even though the Hamiltonian contains just nearest-neighbour interactions, the three-site gate may lead
to better optimizations with respect to the case with two sites (see below).

The SU technique represents a direct generalization of iTEBD to the two-dimensional setting. Starting with some initial state $|\Phi_0\rangle=|\Phi(\{ \Gamma^{(0)} \}, 
\{ \lambda^{(0)} \})\rangle$, we consider the action of a gate on three tensors (e.g., $\Gamma_a^{(0)}$, $\Gamma_b^{(0)}$ and $\Gamma_c^{(0)}$) within a single unit 
cell. The resulting state $|\bar{\Phi}\rangle$ can be exactly expressed as an iPEPS of the same form; however, the auxiliary dimension of the three tensors, as well 
as the weights along the affected bonds ($\lambda_2^{(0)}$ and $\lambda_6^{(0)}$), must be increased. Therefore, we look for a new iPEPS $|\Phi_{\rm loc}\rangle$ with 
new tensors $\Gamma_a^{(1)}$, $\Gamma_b^{(1)}$, and $\Gamma_c^{(1)}$ (while the tensor $\Gamma_d^{(0)}$ is not modified), together with new weights $\lambda_2^{(1)}$ 
and $\lambda_6^{(1)}$, with the original auxiliary dimension $D$. These new tensors are obtained by solving a {\it local} problem that is determined by minimizing the 
distance:
\begin{equation}\label{eq:minSU}
f_{\rm SU} = ||\bar{\Phi}\rangle - |\Phi_{\rm loc}\rangle|,
\end{equation}
which is depicted graphically in Fig.~\ref{fig:distance}. This approach takes a simple form by approximating the environment of the affected sites to be a product of 
weights $\{ \lambda \}$ on the bonds connecting these sites with the rest of the network. This simplified problem is solved by a series of SVDs~\cite{corboz2010}. The last
step consists of replacing these new tensors and weights in all unit cells, which defines the new state $|\Phi_1\rangle$, concluding the SU process for a single 
Trotter gate. The alternation over all possible gates is iterated until convergenece.

The FU optimization shares some aspects with the SU one, namely solving a simple {\it local} problem where the Trotter gate is applied only in a single unit cell. 
Here, the environment is no longer taken to be the trivial product of weights. Given an iPEPS state $|\Psi_0\rangle=|\Psi(a^{(0)},b^{(0)},c^{(0)},d^{(0)})\rangle$, with 
its environment $\{ C^{(0)},T^{(0)} \}$, we apply a single gate on a given position on the lattice acting on three tensors (e.g., $\{ a^{(0)},b^{(0)},c^{(0)} \}$).
As before, this leads to a state $|\bar{\Psi}\rangle$ with tensors having an increased bond dimension along the affected bonds (while the tensor $d^{(0)}$, as well as 
the environment, is not modified). Then, we aim to replace the tensors with an enlarged bond dimension by new tensors $\{ a^{(1)},b^{(1)},c^{(1)} \}$ with the original 
bond dimension $D$. These tensors are given by minimizing the distance:
\begin{equation}\label{eq:minFU}
f_{\rm FU} = ||\Psi_{\rm loc}\rangle - |\bar{\Psi}\rangle|,
\end{equation}
where $|\Psi_{\rm loc}\rangle$ denotes a state where the new tensors are substituted only in the single $2 \times 2$ unit cell, while keeping the same environment 
$\{ C^{(0)},T^{(0)} \}$ (see Fig.~\ref{fig:distance}). The minimization problem is solved by alternating least squares (ALS) as in Ref.~\cite{reza2018}. At each 
step of the ALS, the distance $f_{\rm FU}$ is minimized with respect to a single tensor (out of $a$, $b$, and $c$) while keeping the other ones fixed. The optimized 
tensor is alternated until the convergence of $\Delta f_{\rm FU}$ between two consecutive iterations under the desired threshold $\epsilon$. Typically, we take 
$\epsilon$ between $10^{-7}$ and $10^{-3}$. Finally, the state $|\Psi_1\rangle=|\Psi(a^{(1)},b^{(1)},c^{(1)},d^{(1)})\rangle$ is obtained by replacing the original 
tensors $\{ a^{(0)},b^{(0)},c^{(0)},d^{(0)} \}$ with the new set $\{ a^{(1)},b^{(1)},c^{(1)},d^{(1)}=d^{(0)} \}$ in the entire lattice and by recomputing the 
environment that is compatible with these tensors. Again, this process is iterated until convergence alternating the gates.  

To express the distance, we approximate the environment of the subsystem where the Trotter gate acts. This is the point where SU differs from FU: within SU, the 
environment is taken to be simply a tensor product of weights, thus neglecting most of the correlations in the environment. Every SU step is computationally cheap, 
but the approximation of the distance is very crude. By contrast, within FU, we always approximate the environment using CTM, thus leading to a more accurate distance 
at the expense of the leading computational cost of CTM, i.e., $O(D^{10})$. Still, within this approach the same environment is used for both the old and new tensors. 
The main shortcoming of the procedure is the assumption, that the solution of the local problem (minimizing $f_{\rm FU}$) is also a good solution of the global one, 
i.e., minimizing ${\cal F}=||\Psi_1\rangle - U_{a,b,c} |\Psi_0\rangle|$, where $U_{a,b,c}$ contains all the non-overlapping Trotter gates (here, for the triad $a-b-c$) 
in the infinite lattice (see Fig.~\ref{fig:distance}). 

In general, for any fixed time step $\tau$, the energy generated by FU optimization reaches a minimum and then starts increasing. To have a well defined convergence 
criterion for FU we use an adaptive $\tau$. Should the energy increase after the FU iteration, we go back to the previous state and halve the time step, i.e., 
$\tau \to \tau/2$. The FU optimization is terminated once the time step becomes smaller than $10^{-6}$. Finally, to decrease the computational costs of both SU and FU, 
we use the scheme with reduced tensors where the tensors affected by the action of the Trotter gate are split in two parts: one containing the physical index and the 
auxiliary indices that are involved in the application of the gate and the other one containing all the remaining indices; the latter part is taken to be constant 
and absorbed into the environment~\cite{corboz2010,reza2018}. Moreover, in most of the FU simulations, we do not recompute the environment from scratch after updating the 
tensors; instead, we use the so-called fast FU scheme~\cite{phien2015}, taking only a single iteration of the CTM per applied Trotter gate. Instead, for the evaluation of 
the observables the CTM is always iterated until convergence. All the computations have been performed with \textsc{pi-peps}~\cite{pipeps}, a library for running iPEPS 
simulations built on top of \textsc{itensor}~\cite{itensor}. 

\section{Results}\label{sec:results}

Let us discuss the results of the optimization technique for both the paramagnetic and magnetically ordered phases of the Heisenberg model on coupled two-leg 
ladders of Eq.~(\ref{eq:hamilt}). First of all, it is important to emphasize that, within both the SU and FU techniques, the energy can have a non-monotonic behavior 
along the optimization procedure. Indeed, the minimization problems of Eqs.~(\ref{eq:minSU}) and~(\ref{eq:minFU}) do not necessarily imply that the energy will 
decrease at every step of the evolution. In general, after a relatively short transient in which the energy is rapidly decreasing, a minimum is reached and then 
a slow but inescapable upturn is obtained, no matter how small the imaginary-time discretization is. This is due to the fact that the optimization designed 
within SU or FU does not coincide with a true energy minimization~\cite{corboz2016,vanderstraeten2016}.
\begin{figure}[t]
\includegraphics*[width=\columnwidth]{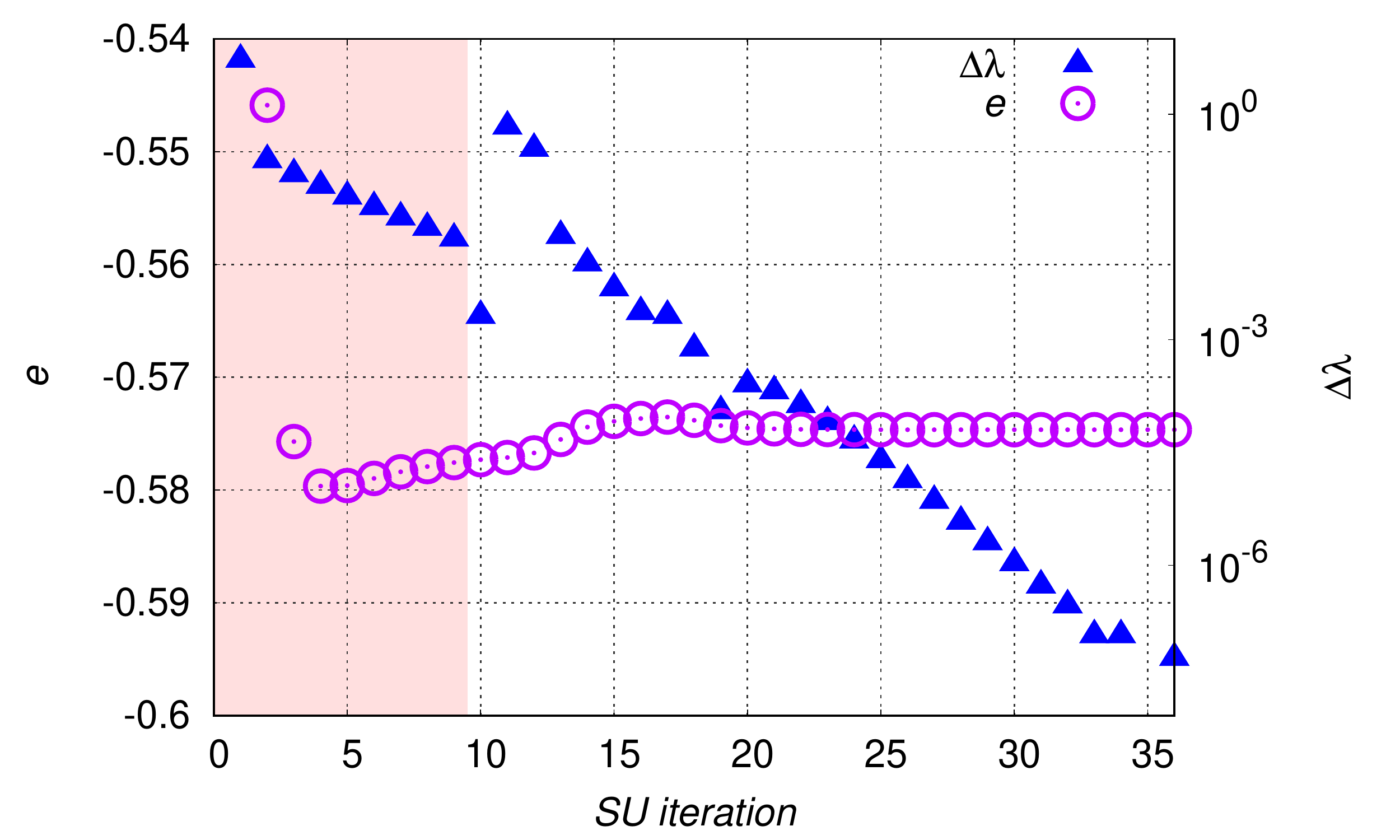}
\caption{\label{fig:suEvsWD} 
Typical SU simulation with $D=4$ for $\alpha=0.2$. The initial tensors reproduce a valence-bond solid with singlets on the strong rungs of the ladders; along the 
optimization three-site gates are used with time step $\tau=0.05$. After each SU iteration (corresponding to the application of 32 gates given by symmetric Trotter-Suzuki 
decomposition) the energy is computed (purple circles, left axis) for the resulting state using CTM with $\chi=96$. The convergence of SU is tracked by the distance 
$\Delta\lambda$ of Eq.~(\ref{eq:deltalam}) (blue triangles, right axis) of weights $\{ \lambda \}$ between consecutive SU iterations. The pink area corresponds to 
states with vanishing magnetization.}
\end{figure}
As an example of this behavior, we report in 
Fig.~\ref{fig:suEvsWD} an optimization performed within SU for $D=4$. Here, we consider $\alpha=0.2$, initializing the tensors in order to have a valence-bond solid, 
in which singlets are formed along the strong rungs of the ladders. Tracing the convergence within SU is often done by observing the change in the weights:
\begin{equation}\label{eq:deltalam}
\Delta \lambda=\sqrt{\sum_{i=1}^{8} \left [ \lambda_i^{(\alpha+1)}-\lambda_i^{(\alpha)} \right ]^2},
\end{equation}
between two subsequent iterations $(\alpha+1)$ and $(\alpha)$, where the weights are always normalized such that the leading weight $\lambda_1=1$. However, while 
$\Delta \lambda$ eventually decreases down to very small values, signaling a converged SU simulation, the energy (computed with full environment by CTM) shows a 
non-monotonic behavior with a clear upturn after a few iterations. In this case, a fixed $\tau=0.05$ is used in order to emphasize the existence of a minimum in the 
energy; by using an adaptive time step, as described at the end of Sec.~\ref{sec:methods}, it would be possible to avoid the rise of the energy, which is otherwise 
inevitable. Most remarkably, even though the exact ground state has a vanishing magnetization $m$ and the initial state has $m=0$, a few steps after the minimum, the 
magnetization becomes finite, spoiling the correct feature of the true ground-state wave function.
\begin{figure}[t]
\includegraphics*[width=0.49\columnwidth]{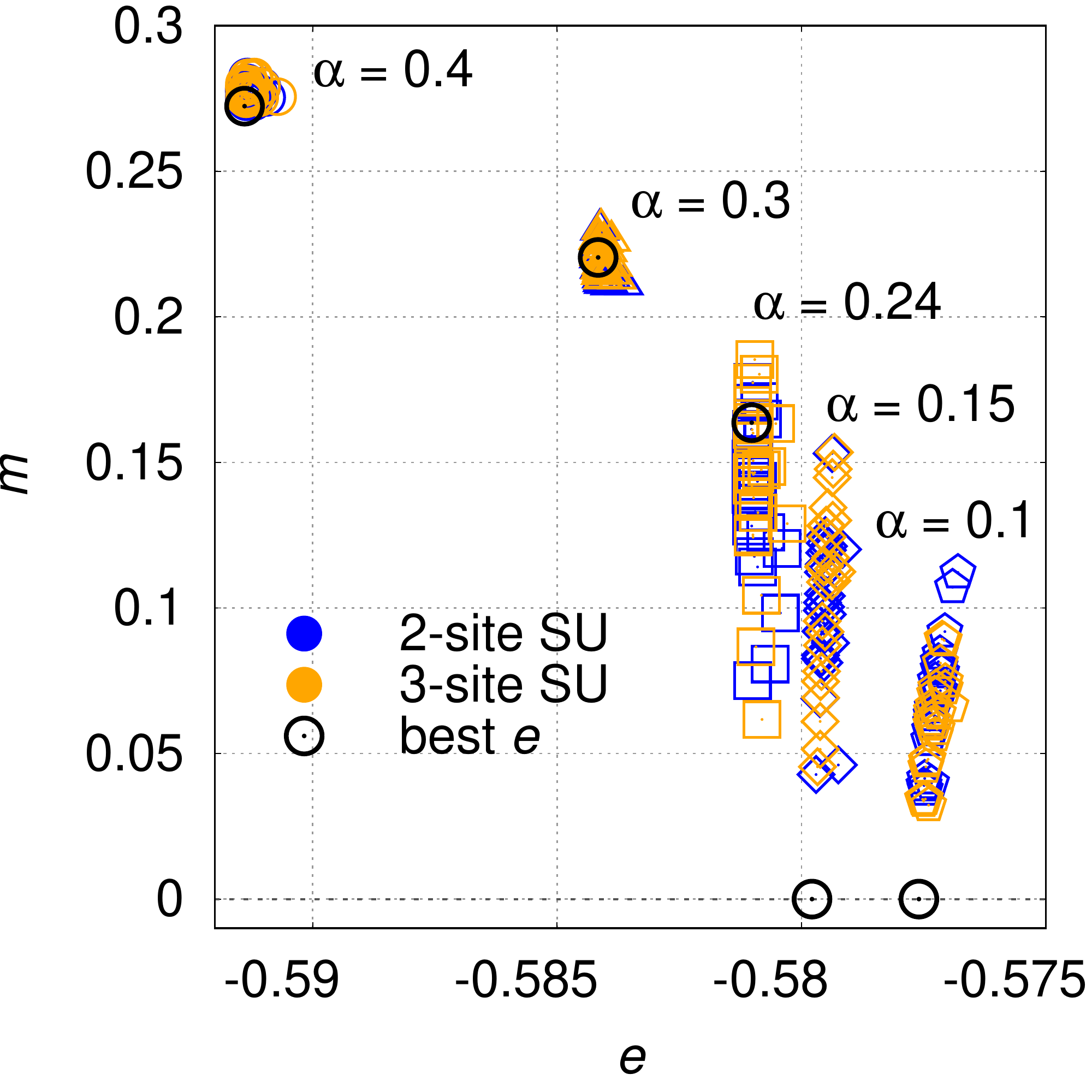}
\includegraphics*[width=0.49\columnwidth]{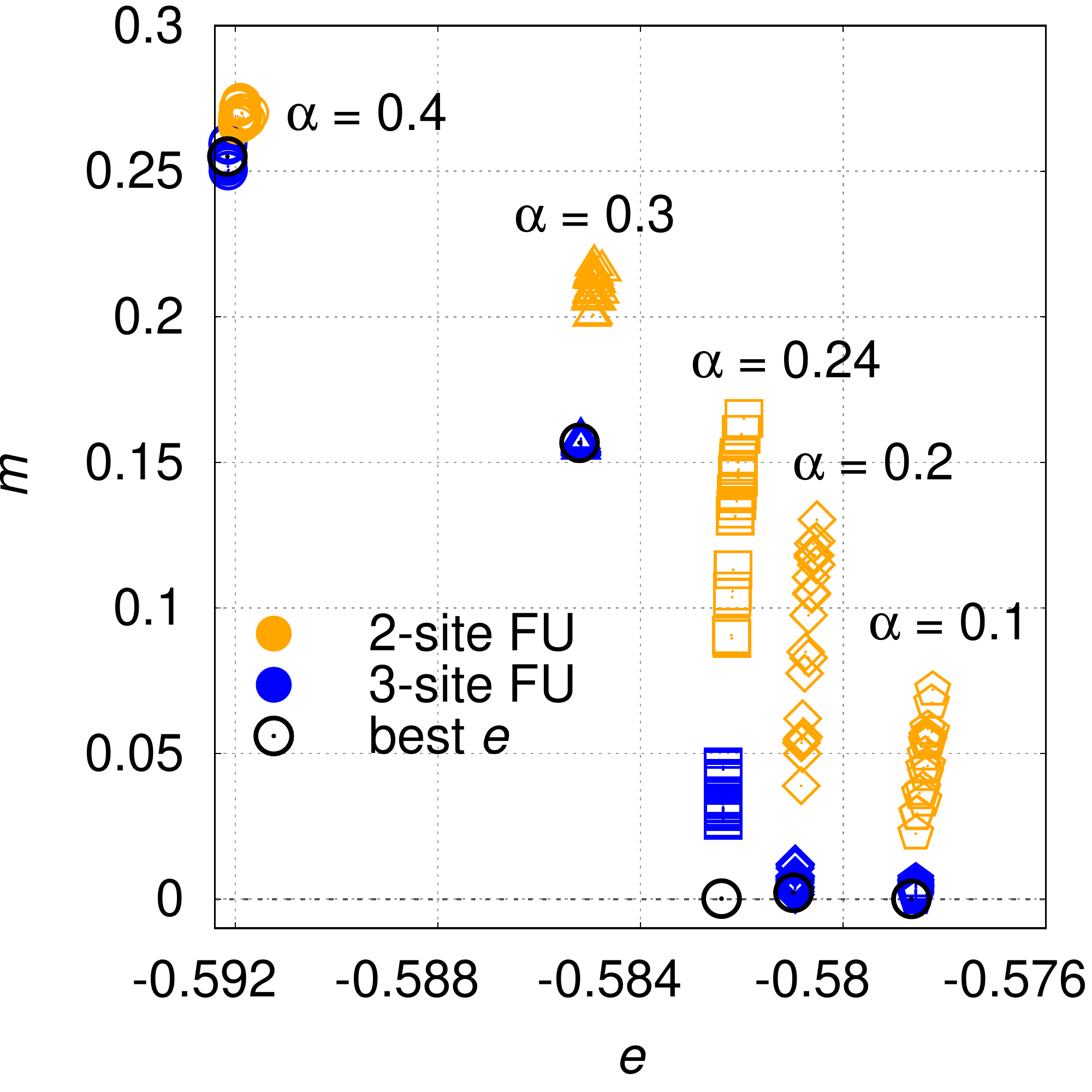}
\caption{\label{fig:2svs3supdate} 
Different optimizations by imaginary-time evolution for $D=4$ with two-site and three-site gates for both SU (left) and FU (right) for a range of values $\alpha$ across the 
transition. The environment bond dimension used in CTM is $\chi=96$. Black circles denote the best energy states for $D=4$ obtained throughout all the simulations within 
SU and FU (see text).}
\end{figure}
Hence, in the spirit of the variational principle, we take the 
lowest-energy state as the ``converged'' one, for which all the other physical properties (i.e., correlation functions) are computed. At the outset, computing the 
energy at every iteration of SU seems to betray its purpose, as a crude but fast way to explore the phase diagram; however, if only states given by converged 
$\Delta \lambda$ are analyzed, the result gives a completely wrong picture with a finite magnetization down to the limit of decoupled ladders.

Now, we would like to stress that both the SU and FU schemes do not always lead to a unique converged state; that is, different starting points may lead to different 
resulting states. In general, this is not a surprising behavior for nonlinear optimization, a case of both SU and FU. Yet for the model of Eq.~(\ref{eq:hamilt}), 
whereas the final energy varies in a relatively small range, other quantities might show considerably stronger variation~\cite{noteeps}. In addition, we find that 
imaginary-time evolutions performed with two- or three-site gates may give distinct results, especially within FU. First of all, we briefly discuss the comparison between 
optimizations done with these two sets of gates for $D=4$, see Fig.~\ref{fig:2svs3supdate}. Within SU, the difference between two-site and three-site gates is small and 
there is no notable advantage in using 3-site gates to perform imaginary-time evolution. Instead, within the FU scheme there is considerable profit in the optimization 
using three-site gates. Two aspects must be emphasized. The first one is that the distribution of the magnetization is much wider in the paramagnetic phase than in the 
antiferromagnetic one for both SU and FU approaches.
\begin{figure}[t]
\includegraphics[width=\columnwidth]{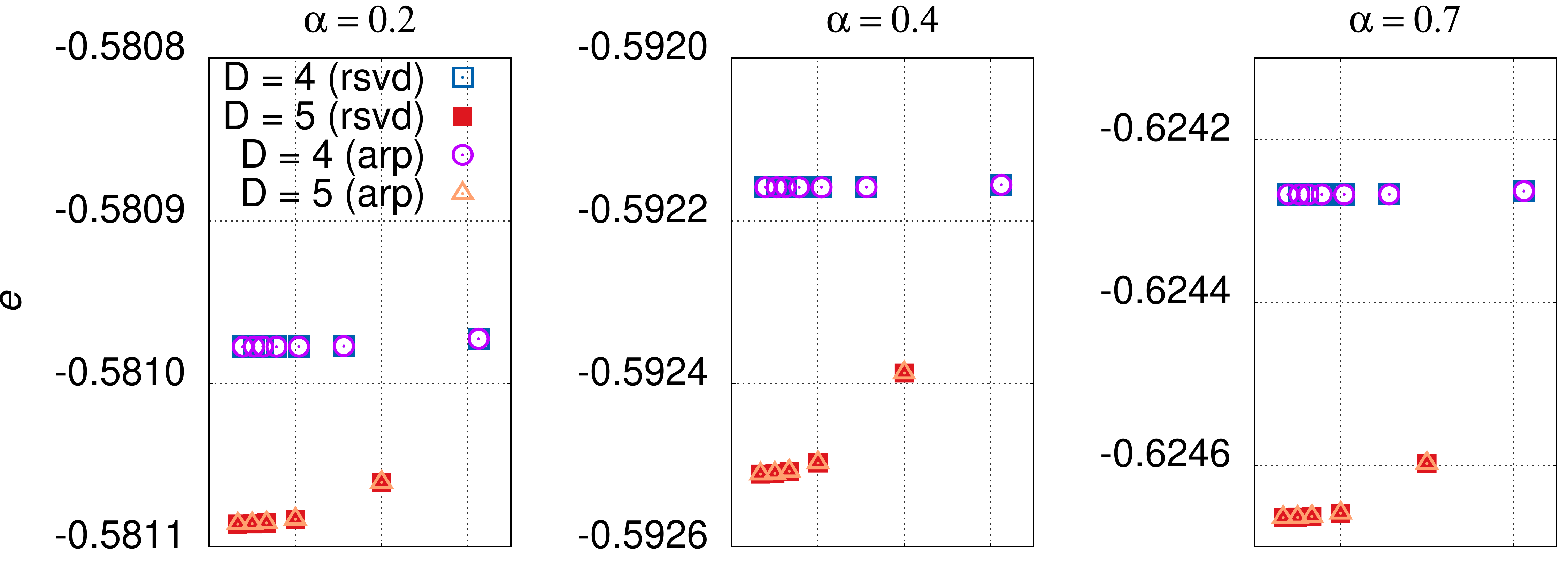}
\includegraphics[width=\columnwidth]{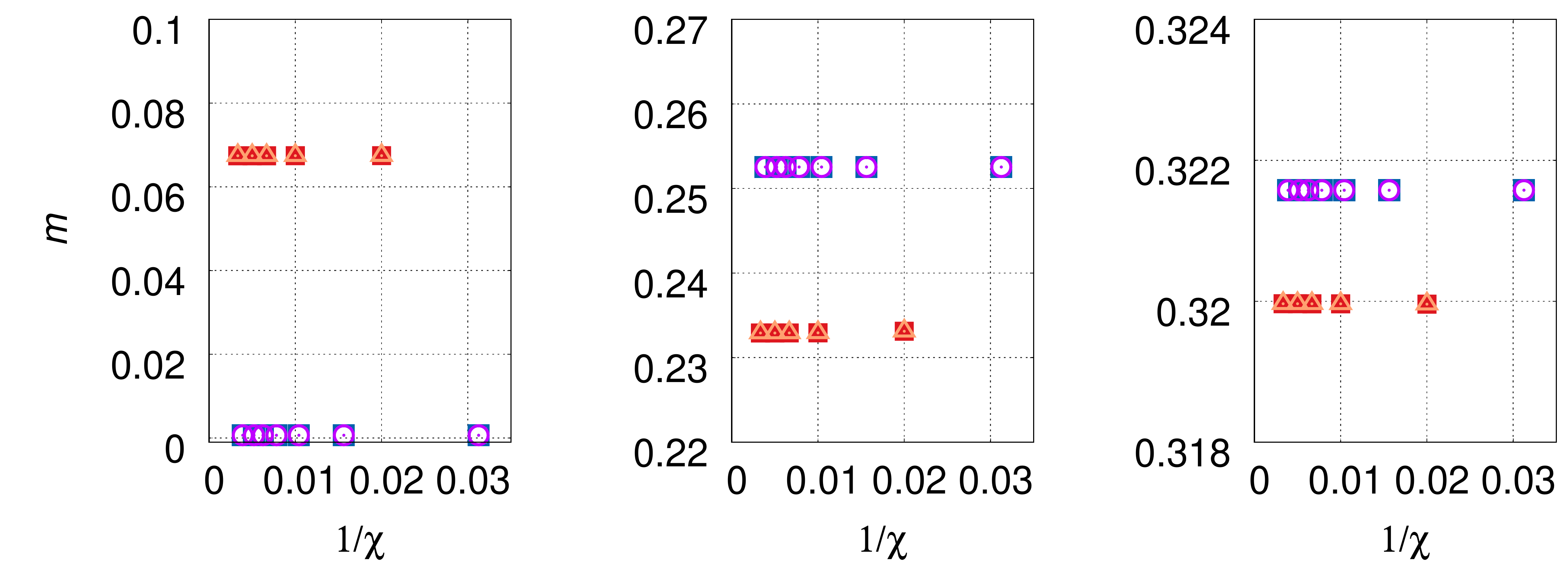}
\caption{\label{fig:chiscaling} 
Scaling of the energy (top panels) and the magnetization (bottom panels) with the environment bond dimension $\chi$ as computed from the best variational states for 
a set of interladder couplings $\alpha$. Results from CTM using randomized SVD (rsvd) or Arnoldi iteration (arp) are found to be in perfect agreement. Magnetization 
is converged for small values of $\chi$ across all cases. Instead, complete convergence of the energy requires higher values of $\chi$ especially close to 
criticality. Nevertheless, even for $\alpha=0.4$ (and $D=5$), very good accuracy can be achieved for a modest value of $\chi=100$, with the difference between 
converged values of energy at $\chi=300$ and $\chi=100$ being less than $1.4 \times 10^{-5}$.}
\end{figure}
\begin{figure}[b]
\includegraphics*[width=0.49\columnwidth]{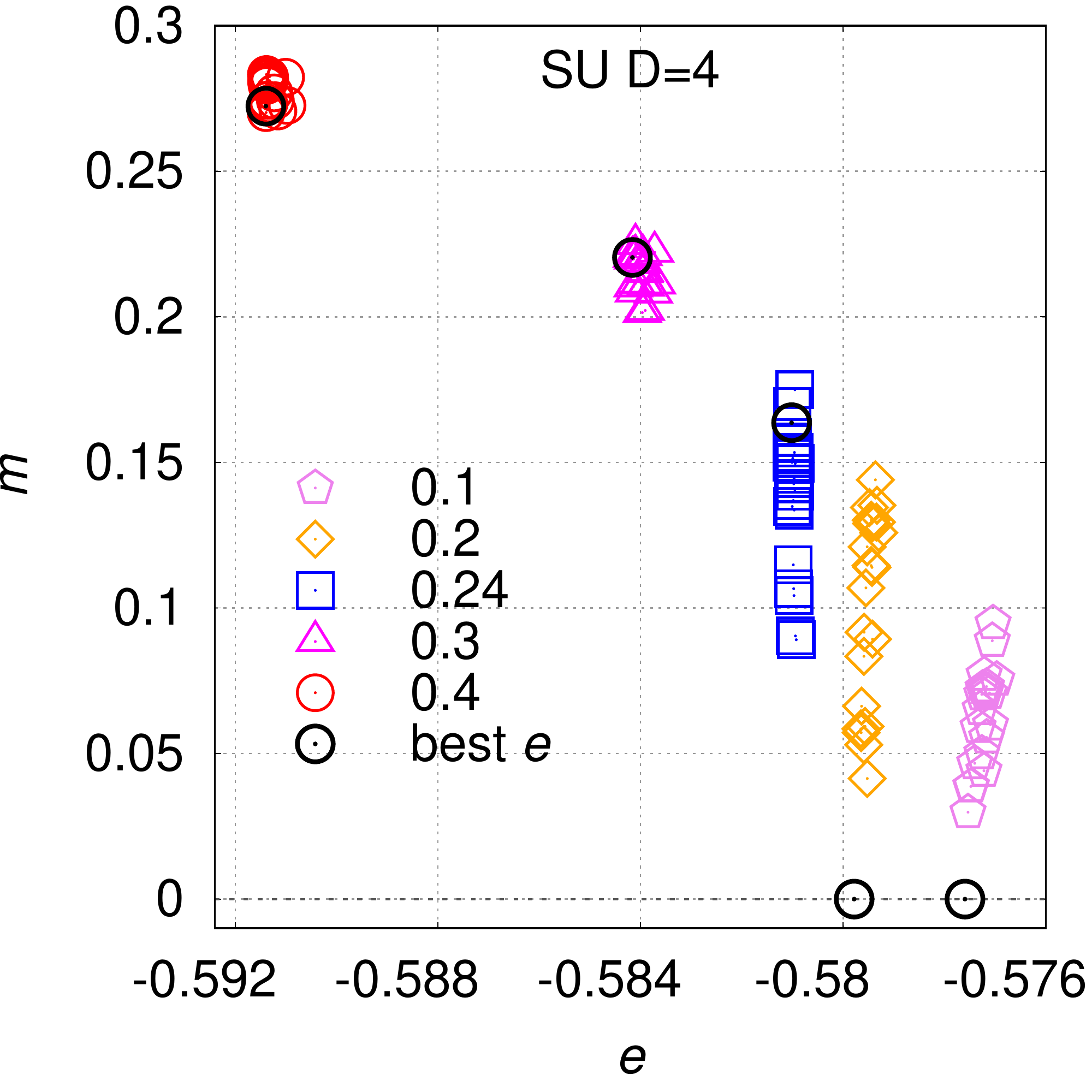}
\includegraphics*[width=0.49\columnwidth]{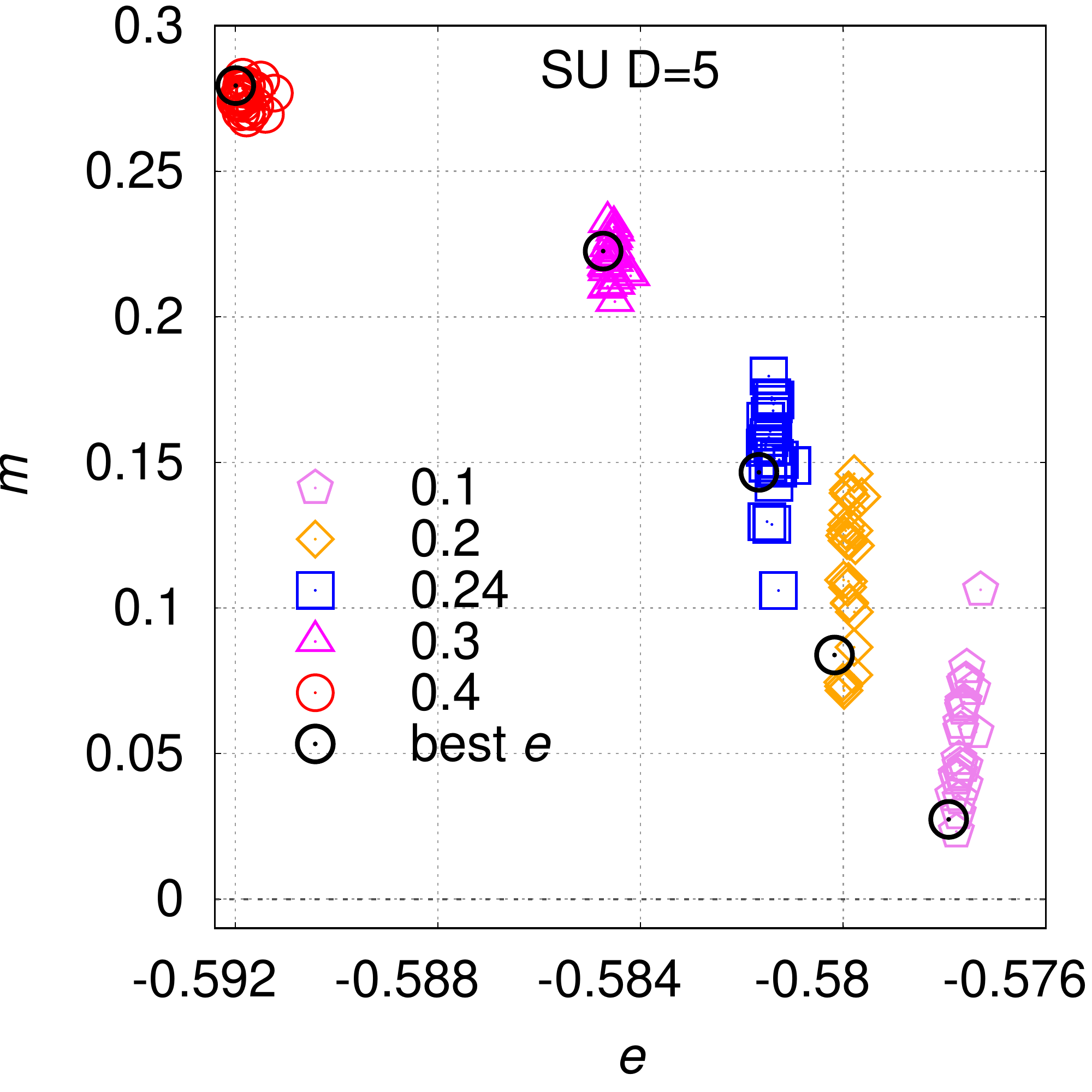}
\caption{\label{fig:d4d5su}
Results for the energy and magnetization as obtained within SU when starting from different (randomized) initial states for different values of the interladder
coupling $\alpha$. The value of the auxiliary bond dimension is $D=4$ (left) and $D=5$ (right). Black circles denote the best energy states within SU obtained throughout 
all the simulations for $D=4$ and $D=5$.}
\end{figure}
Indeed, within the magnetically ordered phase, all the final energies and magnetizations are distributed in a 
very narrow region; most importantly, the fluctuations of $m$ are small with respect to its actual value. By contrast, within the paramagnetic region, it is possible 
to stabilize states with huge variations in $m$, still having tiny energy differences (e.g., of the order of $0.0005J$). We would like to emphasize that the presence 
of large fluctuations in the magnetization persists far away from the critical point, inside the paramagnetic region. 
\begin{figure}[t]
\includegraphics*[width=0.49\columnwidth]{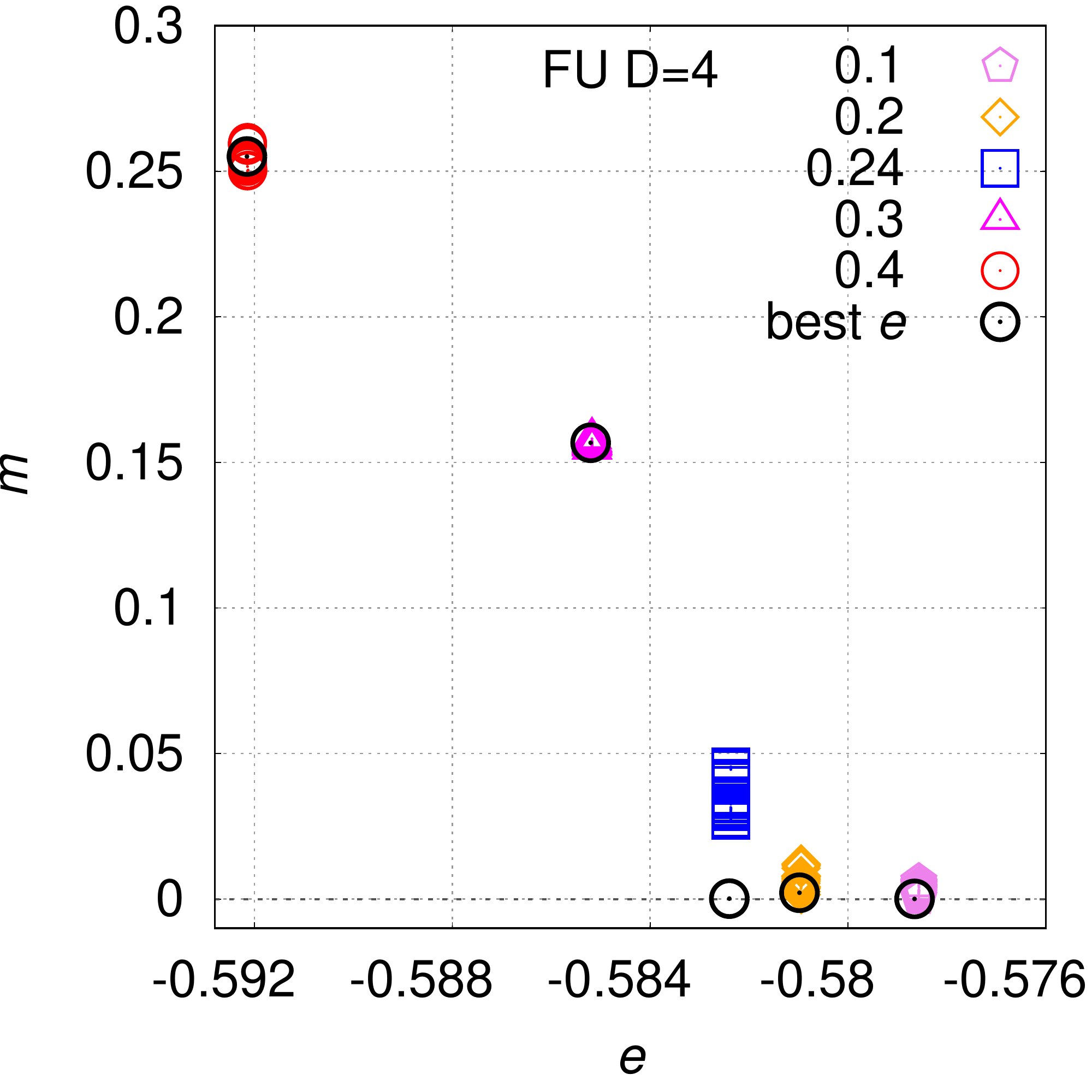}
\includegraphics*[width=0.49\columnwidth]{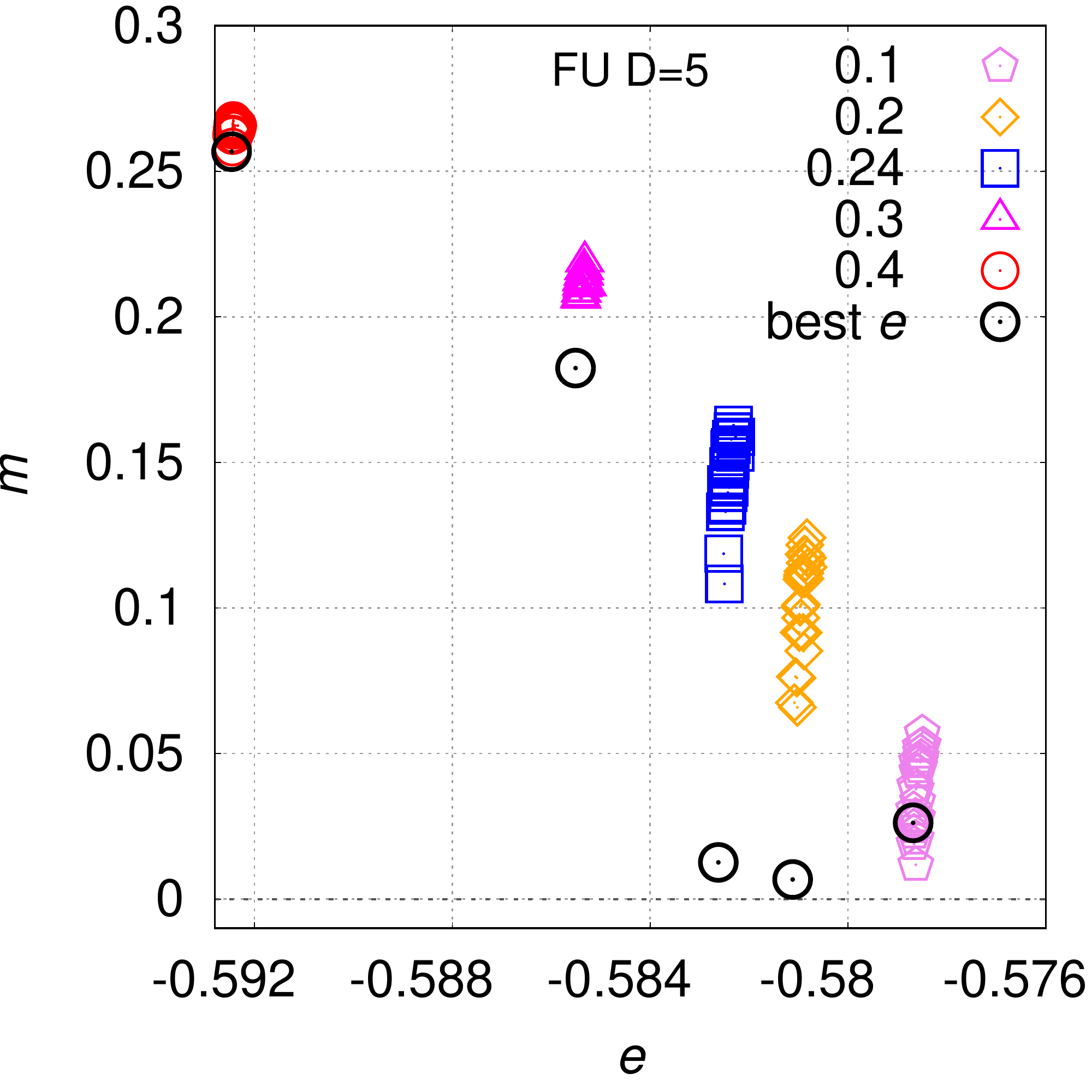}
\caption{\label{fig:d4d5fu}
The same as in Fig.~\ref{fig:d4d5su}, but obtained within FU. Here, the value of $\chi$ that determines the dimension of environment tensors is $96$ for $D=4$ and
$100$ for $D=5$. Black circles denote the best energy states within FU obtained throughout all simulations for $D=4$ and $D=5$ iPEPS (see text).}
\end{figure}
\begin{figure}[b]
\includegraphics*[width=\columnwidth]{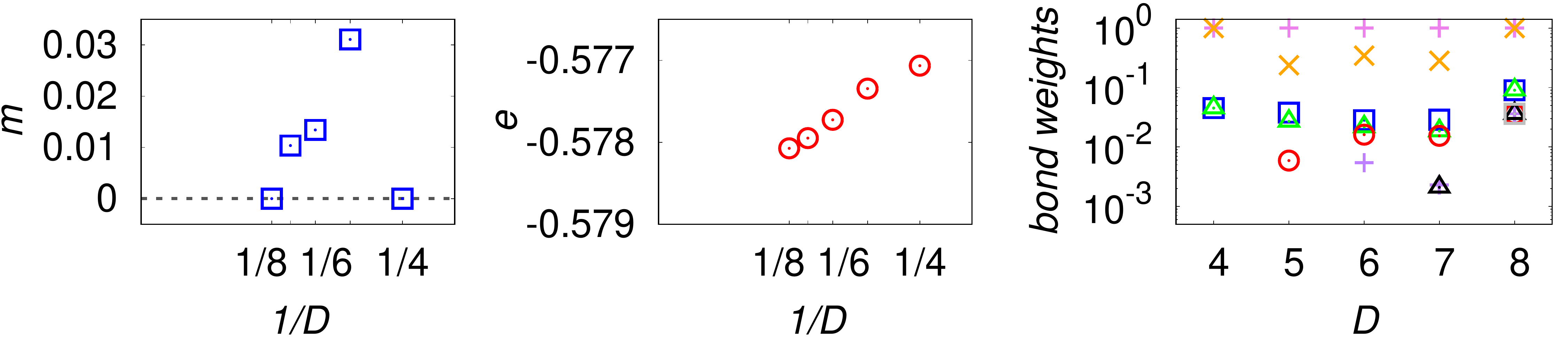}
\includegraphics*[width=\columnwidth]{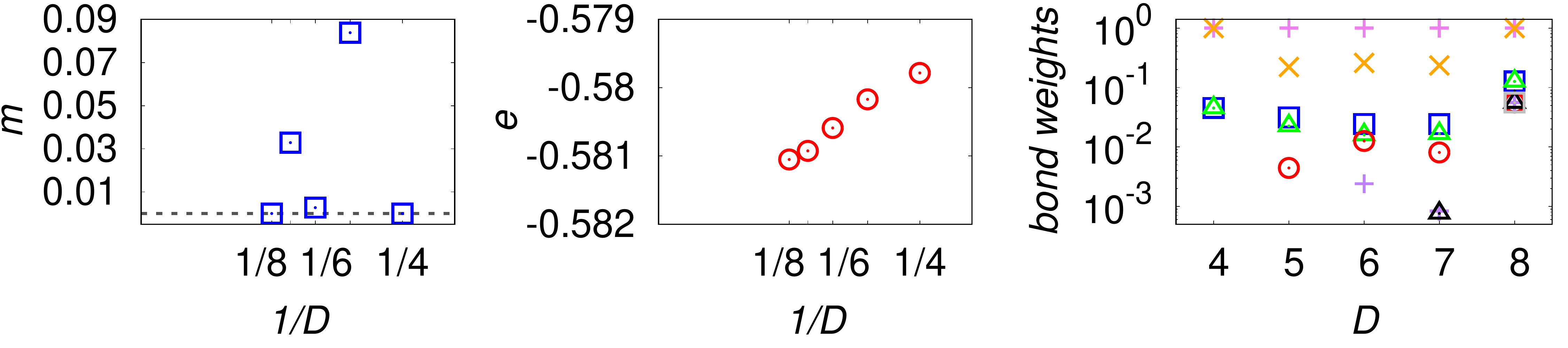}
\includegraphics*[width=\columnwidth]{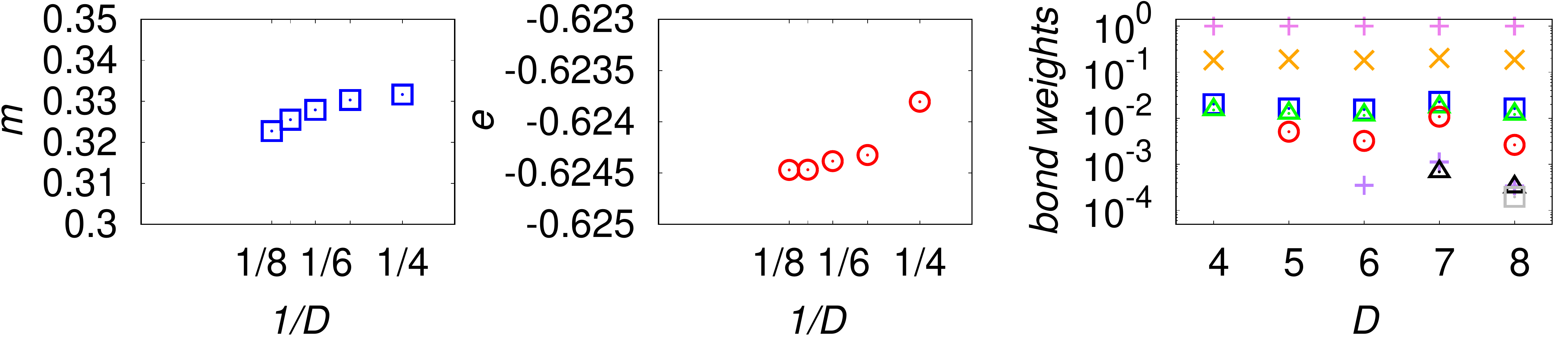}
\caption{\label{fig:spectraSU}
Magnetization, energy, and values of the weights along rungs of the ladders (see text) obtained by SU in the paramagnetic phase with $\alpha=0.05$ (upper panels) and 
$\alpha=0.2$ (middle panels) and in the magnetically ordered phase with $\alpha=0.7$ (lower panels).}
\end{figure}
This aspect is associated with the nature of the 
tensor structure of the wave function and is not related to the presence of a quantum phase transition.
The second aspect, which is by far much more relevant, is that 
a generic optimization that starts from random initial tensors does not give the correct vanishing magnetization within the paramagnetic phase. This is particularly 
true within SU, while the FU scheme highly improves the quality of the results. Still, paramagnetic states are obtained by requiring both a carefully selected initial
state, e.g., valence-bond solids, and a particular value of the auxiliary bond dimension, for example, $D=4$. Let us note that it is possible to induce a finite 
magnetization by breaking SU(2) symmetry in an approximate environment, as observed in Ref.~\cite{poilblanc2017}. Its true vanishing value is then recovered only in the 
limit of $\chi \rightarrow \infty$. However, this is not our case, as we show in Fig.~\ref{fig:chiscaling}. Indeed, the magnetization is well converged for the values 
of environment dimension $\chi$ used.

\begin{figure}[tp]
\includegraphics*[width=0.49\columnwidth]{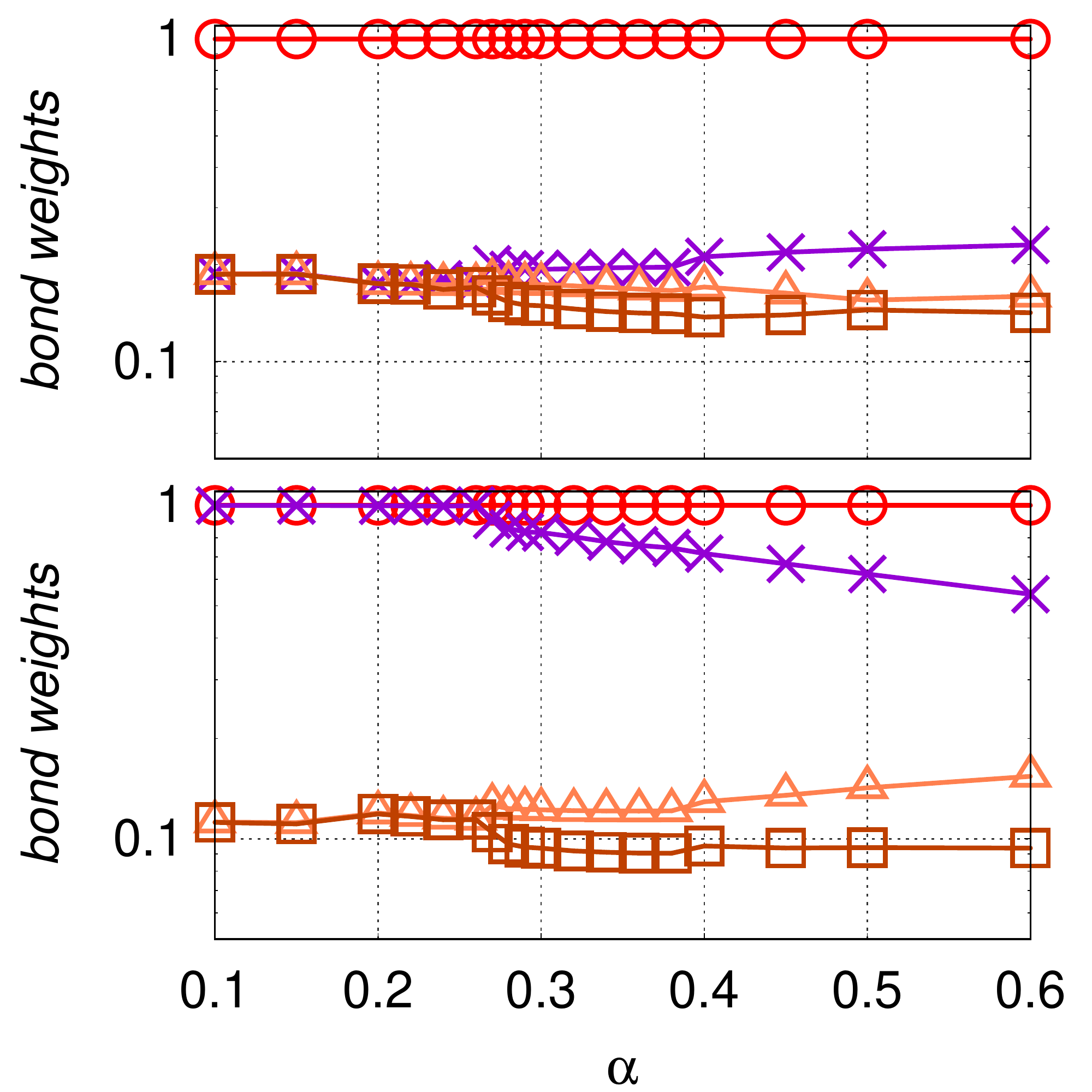}
\includegraphics*[width=0.49\columnwidth]{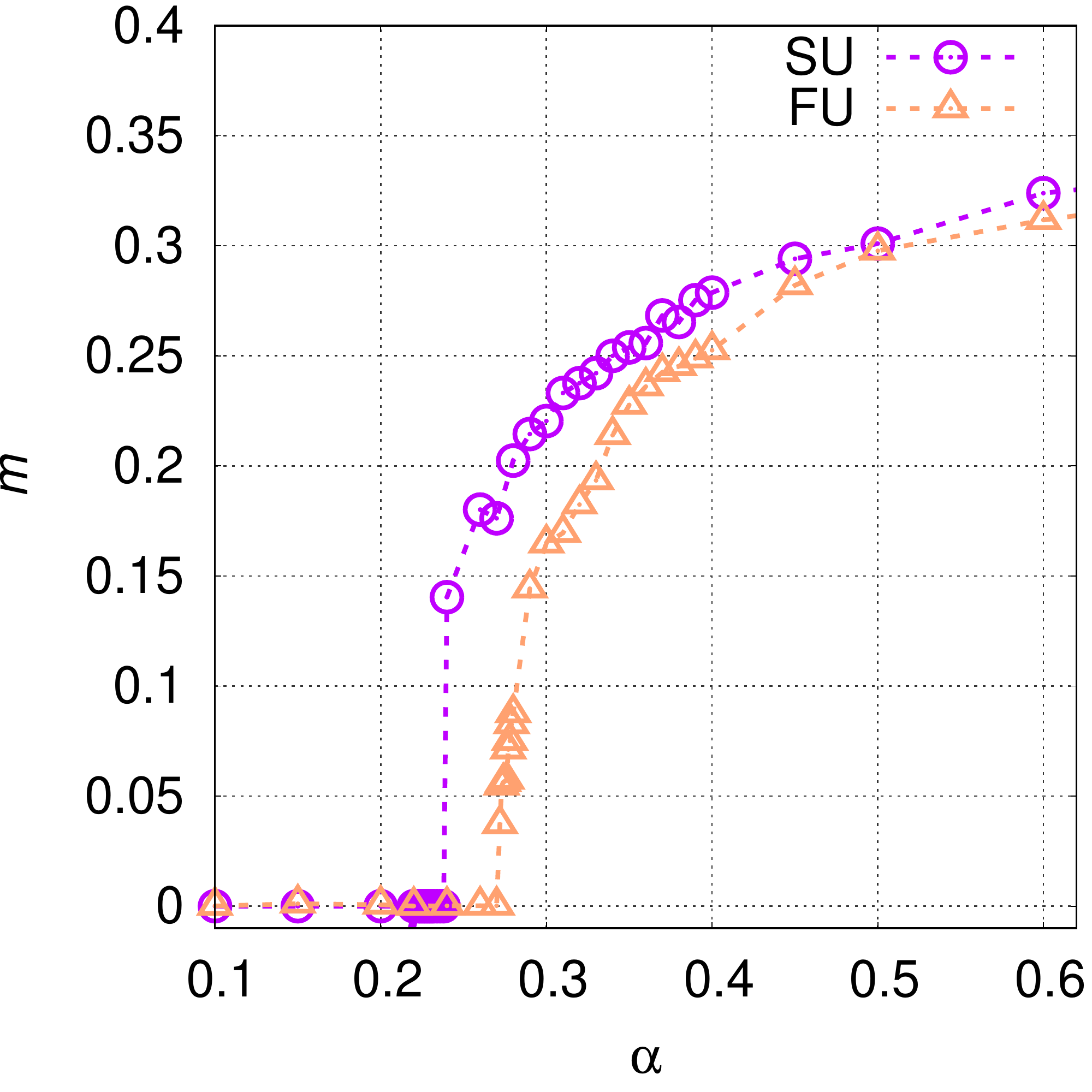}
\caption{\label{fig:phase}
Left: Spectra on selected bonds of $D=4$ optimized by FU across the transition. In the paramagnetic phase, the horizontal bonds (top) running along ladders show degeneracies 
$1$,$3, \dots$, while strong vertical bonds (bottom) show degeneracies $2$, $2, \dots$. Right: Phase diagram of the model of Eq.~(\ref{eq:hamilt}) obtained by using both 
SU and FU with bond dimension $D=4$.}
\end{figure}

For the rest of the paper, we will use three-site gates since, in general, they give better energies with respect to the case with two sites. We now discuss the most important issue of this work, namely, the fact that a paramagnetic state with zero magnetization can be obtained only for selected values of the bond dimension $D$, i.e., the ones 
that do not break the multiplet structure of the tensors. In Fig.~\ref{fig:d4d5su}, we show the outcomes of several SU optimizations for different inter-ladder couplings 
$\alpha$. The cases with $D=4$ and $5$ are presented (the cases with $D=2$ and $3$ give completely unphysical results, with large values of $m$ down to $\alpha=0$, and
therefore will not be discussed here). The results are qualitatively similar when considering the FU technique (see Fig.~\ref{fig:d4d5fu}).

The most remarkable aspect is that the exact result $m=0$ can be obtained only for a few selected values of the bond dimension, i.e., $D=4$ and $D=8$, while 
for all the other values of $D$ the best energy states break the spin SU(2) symmetry and develop finite magnetization. As a consequence, a smooth extrapolation of the 
magnetization with increasing bond dimension $D$ is not always possible, while the energy usually has a very regular behavior (see Fig.~\ref{fig:spectraSU}). In order 
to highlight this feature, we compute the spectrum of the singular values of the matrix that is obtained by contracting the index connecting two neighboring tensors 
and glueing together all the remaining ones of each tensor, thus creating a $2D^3 \times 2D^3$ matrix (where the factor of 2 comes from the physical index). This spectrum is particularly simple within the antiferromagnetic phase, where all singular values are not degenerate (see Fig.~\ref{fig:spectraSU}). In this case, a given
choice of $D$ never spoils the structure of the spectrum, and no appreciable differences are seen in any correlation function. By contrast, within the paramagnetic phase 
a very peculiar multiplet structure appears, which is preserved only for selected values of $D$. Indeed, the spectrum shows degeneracies that depend upon the bond: 
starting from the largest values, we have $1$, $3$, $3$, $1, \dots$ (when contracting along the weak bonds with $\alpha J$ and the strong horizontal bonds along the 
ladder with $J$) and $2$, $2$, $4, \dots$ (when contracting along the strong vertical bonds of the ladder with $J$). Therefore, it is clear that only particular values 
of $D$ can accommodate these multiplet structures (e.g., $D=4$ and $D=8$). In all the other cases, multiplets are broken, which leads to a small residual magnetization. 
Although $m$ can be made relatively small, a faithful extrapolation for $D \to \infty$ is not possible, if not limited to the values of $D$ that give the correct $m=0$ 
result. This outcome poses serious problems whenever we want to describe a paramagnetic (e.g., spin-liquid) phase with a complicated (and not {\it a priori} known) 
multiplet structure. Indeed, it is clear that in this case a blind optimization will very likely lead to a state with a small but finite magnetization, masking the 
existence of a truly quantum paramagnet or spin-liquid phase with vanishing magnetization.

As a consequence of the previous results, the magnetization curve by varying the interladder coupling $\alpha$ is reasonable only for $D=4$ (and $D=8$, not shown),
being finite and smooth (vanishing) for large (small) values of $\alpha$ (see Fig.~\ref{fig:phase}). Still, for this (small) value of the bond dimension the transition 
point is underestimated within SU (i.e., $\alpha \approx 0.24$); in addition, a relatively large jump of the magnetization is observed, in contrast to the exact
behavior where a continuous transition takes places. By emplyoing FU, the critical point shifts towards the correct location (i.e., $\alpha \approx 0.27$), and also 
the jump disappears. Notice that at the quantum critical point the multiplet structure of the tensor is broken, and the ground state develops a finite
magnetization. For other choices of the bond dimension the results are clearly non-physical: for $D=2$ and $3$ a completely smooth curve may be obtained, with $m>0$ 
down to $\alpha=0$. Instead, for $D=5$, $6$, and $7$ it is remarkably hard to work out a smooth curve, and most importantly, finite values of $m$ are still  obtained 
in the paramagnetic regime. This irregular behavior makes it very difficult (if not impossible) to perform an extrapolation for $D \to \infty$. In this context, we
would like to mention that a recently developed finite correlation length scaling (FCLS)~\cite{rader2018,corboz2018} for iPEPS considerably improved the previous estimate 
(extrapolation in $1/D$) of the magnetization in the Heisenberg model (i.e., $\alpha = 1$); here, highly accurate variational states, optimized by gradient methods,
were required. Issues present in the FU simulations pose serious problems for extrapolating the results either in $1/D$ or by using FCLS, especially close to 
criticality.

\vspace*{1em}
\section{Conclusions}\label{sec:concl}

In this paper, we have highlighted a few relevant issues that appear within the iPEPS optimization. First of all, the widely used SU and FU techniques are very 
sensitive to the initial state when applied in a phase with no broken continuous symmetry, giving final states that may have considerably different physical properties 
(e.g., magnetization), while having very close energies. In the example considered here, the $S=1/2$ Heisenberg model on coupled two-leg ladders, this situation is 
particularly evident, since large fluctuations in the magnetization are present within the paramagnetic phase (especially within SU but also within FU). The second 
and the most important aspect, which has not been realized in the past, is the strong dependence of the results on the bond dimension $D$. This feature is intimately 
related to the presence of multiplets in the tensors of symmetric states. In the studied case, the paramagnetic phase is adiabatically connected to a valence-bond
solid (VBS) with singlets along the rungs of the ladders. Both FU and SU  reach the ground state by progressively adding correlations on top of the  initial VBS through 
iterative application of Trotter gates, which themselves have a multiplet structure (the $S_i \cdot S_j$ operator decomposes into a singlet part and a triplet part). 
Without truncation, the multiplets would remain imprinted in the tensors making up the state. However, the multiplet structure is preserved by truncation only for 
specific values of $D$ (e.g., $D=4$ and $8$), giving rise to well-behaved simulations and physically correct variational states. Instead, whenever the value of $D$ 
does not fit the multiplet structure, some breaking mechanism appears, e.g., leading to a finite magnetization and a rough energy landscape. By a similar reasoning, 
we expect multiplets to play a role also in the VBS phase of the $J-Q$ model~\cite{sandvik2007}. In particularly simple models, such as the one that has been considered 
here, it is not hard to find out the exact degeneracy of multiplets and obtain reasonable results, possibly even with a scaling analysis with $D$. In more complicated 
cases (e.g., the frustrated $J_1-J_2$ Heisenberg model on a square or triangular lattice), it might not be easy to work out the degeneracy, possibly leading to 
spurious results, with finite magnetization. In this respect, it is particularly important to impose symmetries in the tensor structure~\cite{jiang2015,mambrini2016} 
and make comparison with unconstrained optimization in order to understand the actual physical properties of highly entangled ground states.
 
\acknowledgements
We acknowledge P. Corboz, P. Czarnik, A. Gendiar, R. Haghshenas, A. Lauchli, and M. Rader for stimulating discussion. We also thank D. Poilblanc, who helped us to  
build the first working PEPS code, and A. Sartori for further development. 

\bibliographystyle{apsrev4-1}

\begin{thebibliography}{99}
\bibitem{noteeps} In general, for a variational state with accuracy $\eta$ on the energy per site, the accuracy on a generic correlation function is $\sqrt{\eta}$. 
   Here, fluctuations in the magnetization are much larger than this estimation.
\end{thebibliography}

\end{document}